\documentclass[11pt]{article}
\usepackage{latexsym}  
\usepackage{amssymb}
\usepackage{graphicx}
\usepackage{amsmath}

\newcommand{\be}{\begin{equation}}
\newcommand{\ee}{\end{equation}}
\newcommand{\bea}{\begin{eqnarray}}
\newcommand{\eea}{\end{eqnarray}}
\newcommand{\bref}[1]{(\ref{#1})}
\begin{document}
\begin{titlepage}
\begin{flushright}
\today
\end{flushright}
\vspace{4\baselineskip}
\begin{center}
{\Large\bf  Universality of Quark-Lepton  Mass Matrix.}
\end{center}
\vspace{1cm}
\begin{center}
{\large Takeshi Fukuyama$^{a,}$
\footnote{E-mail:fukuyama@se.ritsumei.ac.jp}}
and
{\large Hiroyuki Nishiura$^{b,}$
\footnote{E-mail:nishiura@is.oit.ac.jp}}
\end{center}
\vspace{0.2cm}
\begin{center}
${}^{a}$ {\small \it Department of Physics and R-GIRO, Ritsumeikan University,
Kusatsu, Shiga,525-8577, Japan}\\[.2cm]

${}^{b} $ {\small \it Faculty of Information Science and Technology, 
Osaka Institute of Technology,\\ Hirakata, Osaka 573-0196, Japan}

\vskip 10mm
\end{center}
\vskip 10mm
\begin{abstract}
The recently observed lepton mixing angle $\theta_{13}$ of the MNS mixing matrix is well incorporated in a universal mixing hypothesis between quark and lepton sectors.
This hypothesis asserts that, in the charged lepton diagonal base, all other mass matrices for up- and down-type quarks and light neutrinos 
are diagonalized by the same unitary matrix except for the phase elements. It is expressed as $V_{CKM}= U_{MNS}(\delta^\prime)^\dagger P U_{MNS}(\delta)$ for quark mixing matrix $V_{CKM}$ and lepton mixing matrix $U_{MNS}(\delta)$ in the phenomenological level. Here $P$ is a diagonal phase mass matrix. 
$\delta^\prime$ is a slightly different phase parameter from the Dirac CP violating phase $\delta=1.1\pi$ (best fit) in the MNS lepton mixing matrix.

\end{abstract}
\end{titlepage}
The most drastic difference between quark and lepton mixing matrices is summarised as follows: the Cabbibo-Kobayashi-Maskawa (CKM) quark mixing matrix is almost diagonal, whereas the Maki-Nakagawa-Sakata (MNS) lepton mixing matrix is almost maximally mixed.
In this letter, we consider a hypothesis which explains this character very naturally.
When we say some universality ranging over quarks and leptons, we should consider such property in GUT framework in which the quarks and leptons belong to the same multiplet.

For SO(10) GUT, all the quarks and leptons in the standard model (SM) and right-handed neutrinos $\nu_R$ are involved in a single {\bf 16}-plet. If we consider the mass relations in SO(10) framework, the rebasing is performed on the {\bf 16}-plet and not independently on the respective quark and lepton SU(2)$_L$ doublets 
$Q_L,~L_L$, and right-handed singlets $~u_R,~d_R,~e_R,~\nu_R$ unlike at electro-weak (EW) scale. 
The subsequent arguments are indebted to this fact but not to the detail of the model.

Based on the above arguments, we can unitarily rotate {\bf 16}-plet to diagonalize charged lepton mass matrix. 
Hence the MNS matrix ($\equiv U_{MNS}$) diagonalizes light neutrino mass matrix.  Our universal hypothesis is to assume that mass matrices for up and down-type quarks are diagonalized by the same mixing matrix $U_{MNS}$. 
Namely the unitary matrices $U_u$ and $U_d$ diagonalizing the mass matrix for up- and down-quarks, respectively, are the same matrix as $U_{MNS}$ up to diagonal phase matrix $P$,
\be
U_u=P^\dagger U_{MNS},~~U_d= U_{MNS}.
\label{proposition}
\ee
So the CKM quark mixing matrix  ($\equiv V_{CKM}$) is represented by
\be
V_{CKM}= U_{MNS}^\dagger P U_{MNS},
\label{CKM}
\ee
where
\be
P\equiv \left(
\begin{array}{ccc}
e^{i\phi_1}&0&0\\
0&e^{i\phi_2}&0\\
0&0&1\\
\end{array}
\right).\label{P}
\ee
In the previous paper \cite{N-F}, we adopted this hypothesis 
and predicted the lepton mixing angle $\theta_{13}$ before the experimental discovery of it 
by using special form of $U_{MNS}$ and the observed CKM matrix. 
Unfortunately the predicted value ($0.036<s_{13}<0.048$) is too small for the observed value $\sqrt{0.024}=0.155$ \cite{Daya-Bay}. 
So in this letter, we show that the modified hypothesis, with using observed $U_{MNS}$, 
satisfies all the observed data of the MNS and CKM and explains the implications of the modification.

Since neutrino oscillation experiment is wholly insensitive to the Majorana CP violating phases, 
$U_{MNS}$ may be in general written using only the Dirac CP violating phase in the standard form given by 
\begin{equation}
U_{MNS}(\delta)=
\left(
\begin{array}{ccc}
c_{13}c_{12},&c_{13}s_{12},& s_{13}e^{-i\delta}\\
-c_{23}s_{12}-s_{23}c_{12}s_{13}e^{i\delta},&c_{23}c_{12}-s_{23}s_{12}s_{13}e^{i\delta},&s_{23}c_{13}\\
s_{23}s_{12}-c_{23}c_{12}s_{13}e^{i\delta},&-s_{23}c_{12}-c_{23}s_{12}s_{13}e^{i\delta},&c_{23}c_{13}
\end{array}
\right) 
\label{mixing}
\end{equation}
for the analysis of the neutrino oscillation data. Here $s_{13}=\mbox{sin}\theta_{13},~c_{13}=\mbox{cos}\theta_{13}$ etc. 
In the following discussions, we adopt the global best fit values by Fogli et al. \cite{Fogli} 
for the mixing angles and the CP violating phases in (\ref{mixing}), which are given by
\bea
s_{12}=\sqrt{0.31},~s_{23}=\sqrt{0.39},~s_{13}=\sqrt{0.024},~
\delta=1.1 \pi.
\eea

Substituting these values into \bref{CKM} and using the following four observed values\cite{PDG} of the CKM matrix elements,
\begin{eqnarray}
|(V_{CKM})_{us}|&=&0.2252 \pm 0.0009,\label{CKMus}\\
|(V_{CKM})_{cb}|&=&0.0409 \pm 0.0011,\label{CKMcb}\\
|(V_{CKM})_{ub}|&=&0.00415 \pm 0.00049,\label{CKMub}\\
|(V_{CKM})_{td}|&=&0.0084 \pm  0.0006,\label{CKMtd}
\end{eqnarray}  
let us fit our relation (\ref{mixing}) by using two free parameters $\phi_1$ and $\phi_2$.
This is equivalent to the whole data fittings for the $V_{CKM}$.
Unfortunately we find that we have no solution which satisfies all four constraints on $|V_{CKM}|$.
It may come from a following reason.  So far we have considered the data of MNS and CKM at EW energy level.
If the universal hypothesis is valid at a GUT level, we must run down the relation to the SM scale 
by renormalization group equation (RGE).
Its effect is not so large but may affect \bref{CKM}.

The RGE effect on CKM was considered, for example,  in \cite{Fuku}\cite{R-S} for  SO(10) GUT.  
In terms of the Wolfenstein parameters, RGE effects of $A, ~\eta$, the others \{$\rho,~\lambda$\} are relatively large, small, negligible, respectively in general \cite{R-S}.
However, in large tan$\beta$ case like SO(10) GUT, RGE effect is rather restricted in CKM phase.

Therefore, at the EW energy scale, let us replace \bref{proposition} with
\be
U_u=P^\dagger U_{MNS}(\delta^\prime),~~U_d= U_{MNS}(\delta),
\label{proposition2}
\ee
as an improved hypothesis which takes account of the difference on the RGE effect between the up- and down-type quarks.
Here $U_{MNS}(\delta^\prime)$ is defined only by replacing $\delta=1.1\pi$ in $U_{MNS}(\delta)$ 
by the free phase parameter $\delta'$. Since RGE effect is small, the value of the $\delta'$ must be close to $\delta=1.1\pi$. 
The mixing angles in the $U_{MNS}(\delta^\prime)$ are assumed to be the same as those in $U_{MNS}$.

Thus we have improved relation between $V_{CKM}$ and $U_{MNS}$ as
\be
V_{CKM}= U_{MNS}(\delta^\prime)^\dagger P U_{MNS}(\delta).
\label{CKM_improved}
\ee
The diagonal phase matrix $P$ is given by (\ref{P}).

We now search for values of three free parameters $\phi_1$, $~\phi_2$, and $~\delta'$ in order for the relation \bref{CKM_improved} to be consistent with the observed CKM given by \bref{CKMus} - \bref{CKMtd}.  
For this purpose, we draw the allowed region in the $\phi_1$-$\phi_2$ plane from  \bref{CKM_improved} with the observed CKM for a each given value of $~\delta'$ in the range $0\leq \delta' \leq 2\pi$. 
We find that a consistent relation  is realized only for the case of  $\delta'\simeq \pi$. 
In Figure 1, by taking $\delta'\simeq \pi$, we show the allowed region in the $\phi_1$-$\phi_2$ plane 
which is obtained from  \bref{CKM_improved} with the observed CKM given by \bref{CKMus} - \bref{CKMtd}. 
Thus we obtain a consistent set of parameters such that 
\be
\delta^\prime\simeq \pi,\quad \phi_1\simeq 26.3^\circ, \quad \phi_2\simeq -3.8^\circ, 
\label{parameter_set}
\ee 
as depicted in Figure 1. 
In Figure 2, by taking $\phi_1= 26.3^\circ$ and $\phi_2 = -3.8^\circ$ and by treating $\delta^\prime$ as a free parameter too, we show the allowed region in the $\delta$-$\delta^\prime$ plane obtained from the observed CKM. 
As seen from Figure 2, we find that the consistent relation  \bref{CKM_improved} is realized only for $\delta^\prime\simeq \pi$ and $\delta\simeq 1.1 \pi$.  

\begin{figure}[t]
\centering
\includegraphics[width=0.49\textwidth]{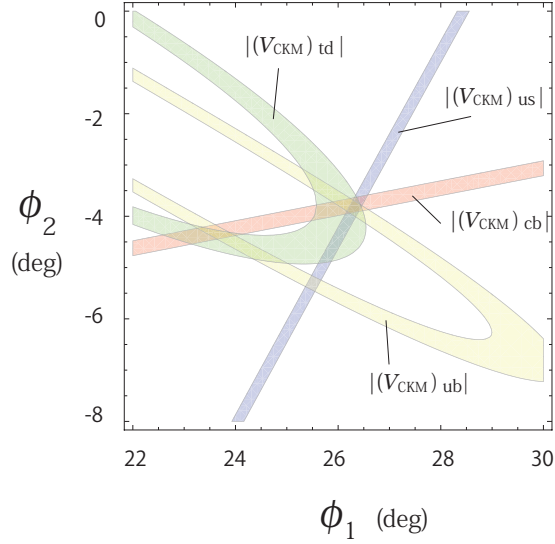}
\caption{Allowed region in the $\phi_1-\phi_2$ plane 
which is consistent with the experimental data of the CKM quark mixing matrix elements for $\delta'=\pi$. 
The shaded areas are allowed, which are obtained from 
the experimental data of $|(V_{CKM})_{us}|$, $|(V_{CKM})_{cb}|$, $|(V_{CKM})_{ub}|$, and $|(V_{CKM})_{td}|$ 
 given by \bref{CKMus} - \bref{CKMtd}.
The overlapping region of them is consistent parameter region with the observed $V_{CKM}$. }
\label{fig1}
\end{figure}

\begin{figure}[t]
\centering
\includegraphics[width=0.49\textwidth]{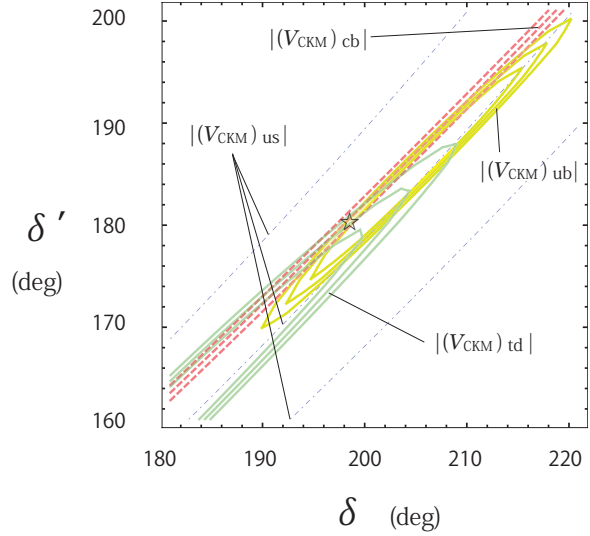}
\caption{Contour plots in the $\delta-\delta^\prime$ plane 
of the observed CKM elements for the case of  $\phi_1= 26.3^\circ$ and $\phi_2 = -3.8^\circ$.  
Contour curves for center, lower, and upper values of $|(V_{CKM})_{us}|$, $|(V_{CKM})_{cb}|$, $|(V_{CKM})_{ub}|$, and $|(V_{CKM})_{td}|$ given by \bref{CKMus} - \bref{CKMtd} as functions of $\delta$ and $\delta^\prime$ are drawn by dot-dashed, dashed, light solid(yellow), and dark solid(green) curves, respectively. 
All the experimental data on CKM are satisfied at the point indicated by the star ($\star$) in the $\delta-\delta^\prime$ plane.}
\label{fig2}
\end{figure}

Taking the parameter set given in \bref{parameter_set} 
( $\delta'= \pi$, $\phi_1= 26.3^\circ$, and $\phi_2= -3.8^\circ$), 
we obtain the following numerical values for $V_{CKM}$ from \bref{CKM_improved}: 
\be
V_{CKM}=\left(
\begin{array}{ccc}
 0.9286\, +0.2950 i & -0.04711+0.2199 i & 0.00181\, -0.00386 i \\
 -0.04556\,+0.2201 i & 0.9681\, +0.1037 i & -0.000061\,-0.04103 i \\
 0.00844\, -0.00242 i & 0.00438\, -0.04007 i & 0.9989\, -0.0218 i \\
\end{array}
\right).
\ee
This $V_{CKM}$ predicts 
\begin{eqnarray}
|(V_{CKM})_{us}|&=&0.2249,\\
|(V_{CKM})_{cb}|&=&0.0410,\\
|(V_{CKM})_{ub}|&=&0.00426,\\
|(V_{CKM})_{td}|&=&0.00878,\label{prediction1}
\end{eqnarray}  
for CKM matrix elements and 
\be
\delta_{q}=70.7^\circ
\ee
for the Dirac CP violating phase in the standard representation of $V_{CKM}$. 
The predicted values are well consistent with the observed data.

Now let us consider the implication of the above arguments and results. 
We have discussed the universal mixing hypothesis and obtained the phenomenological relation 
between the NMS and CKM mixing matrices given by \bref{CKM_improved}.
 In the preceding analysis we adopted the global best fit of \cite{Fogli}.
If we adopt the other global best fit for the NMS mixing angles by Forero et al. \cite{Forero} , we have also solution with different $\phi_i$ but with the same $\delta=1.1\pi$ and $\delta^\prime=\pi$.
So $\delta^\prime=\pi$ seems to have an essential meaning. 
It may indicate that the mass matrices for quarks are real symmetric at the GUT scale 
and CP violating phases are induced by the RGE effect in addition to the diagonal phase matrix P.

The conventional quark-lepton complementarity \cite{Smirnov} claims that the quark- and lepton- mixing angles $\theta_{ij}^q$ and $\theta_{ij}^\ell$ satisfy the relation such as
$\theta_{12}^q+\theta_{12}^\ell=\frac{\pi}{4}$ or $~\theta_{23}^q+\theta_{23}^\ell=\frac{\pi}{4}$,  
which is valid only in the standard representation \cite{Zhang}.  Whereas our
\bref{CKM_improved} is independent of the representation. 
Our phenomenological relation (\ref{CKM_improved}) is a new quark-lepton complementarity relation 
expressed in terms of the mixing matrices and will offer us a suggestive hint 
for building a unified mass matrix model for quarks and leptons.

\section*{acknowledgement}
The work of T.F.\ is supported in part by the Grant-in-Aid for Science Research
from the Ministry of Education, Science and Culture of Japan
(No.~020540282).


\begin{thebibliography}{99}
\bibitem{N-F}
H. Nishiura and T. Fukuyama, Mod. Phys. Lett. {\bf A26}, 661 (2011).
\bibitem{Daya-Bay}
F. P. An {\it et al.}iDaya-Bay Collaboration), arXiv:1203.1669.
\bibitem{Fogli}
G. L. Fogli {\it et al.}, Phys. Rev. {\bf D86}, 013012 (2012) [arXiv:1205.5254].
\bibitem{PDG}
J. Beringer {\it et al.} (Particle Data Group), Phys. Rev. {\bf D86}, 010001 (2012). 
\bibitem{Fuku}
T. Fukuyama and N. Okada, JHEP {\bf 0211}, 011 (2002).
\bibitem{R-S}
G. G. Ross and M. Serna, Phys. Lett. {\bf B664}, 97 (2008).
\bibitem{Forero}
D. V. Forero, M. Tortola and J. W. F. Valle, arXiv:1205.4018. 
\bibitem{Smirnov}
H. Minakata and A. Y. Smirnov, Phys. Rev. {\bf D70}, 073009 (2004); 
S. Antusch, S. F. King and R. N. Mohapatra, Phys. Lett. {\bf B618}, 150 (2005).
\bibitem{Zhang}
Y. Zhang, X. Zhang and B. Q. Ma, arXiv:1211.3198.

\end{thebibliography}
\end{document}